Title

Particle deposition on the saturnian satellites from ephemeral cryovolcanism on Enceladus

Authors

Naoyuki Hirata [a], Hideaki Miyamoto [a, b, ]*, and Adam P. Showman [c]

Authors' affilication

a The University Museum, The University of Tokyo, Hongo, Tokyo 113-0033, Japan.
b Planetary Science Institute, 1700E Ft. Lowell, Suite 106, Tucson, AZ 85719, USA
c Department of Planetary Sciences, Lunar and Planetary Laboratory, University of Arizona, Tucson, AZ 85721, USA
* Corresponding author E-mail address: hm@um.u-tokyo.ac.jp

Abstract

The geologically active south pole of Enceladus generates a plume of micron-sized particles, which likely form Saturn's tenuous E-ring extending from the orbit of Mimas to Titan. Interactions between these particles and satellites have been suggested, though only as very thin surficial phenomena. We scrutinize high-resolution images with a newly developed numerical shape model of Helene and find that the leading hemisphere of Helene is covered by thick deposits of E-ring particles, which occasionally collapse to form gully-like depressions. The depths of the resultant gullies and near-absence of small craters on the leading hemisphere indicate that the deposit is tens to hundreds of meters thick. The ages of the deposits are less than several tens of My, which coincides well with similar deposits found on Telesto and Calypso. Our findings as well as previous theoretical work collectively indicate that the cryovolcanic activity currently occurring on Enceladus is ephemeral.

# 1 Introduction

Mid-sized satellites in the E-ring system, such as Tethys and Dione (co-orbital moon of Helene), show almost bimodal distributions of albedo [Buratti et al., 1990; Verbiscer and Veverka, 1992] and visual and infrared spectra [Pitman et al., 2010]. For example, the leading hemisphere of Dione at visible wavelengths is 1.8 times brighter than the trailing hemisphere [Jaumann et al., 2009], which has been suggested to result from differential accumulation of E-ring material on its surface [Ostro et al., 2010; Verbiscer et al., 2007] and/or by bombardment of corotational plasma and energetic electrons [Schenk et al., 2011]. However, no depositional features have been reported on satellites in the E-ring region except for those on Enceladus (e.g., buried craters) [Kirchoff and Schenk, 2009]. We thoroughly examine all high-resolution images obtained by the Cassini spacecraft through November 2013 and find that possible depositional features ubiquitously exist on small satellites in the E-ring region. The best examples are found on Helene because it is the most extensively imaged small satellite in the E-ring.

# 2 Geological Study of Helene

Like most saturnian satellites, Helene has numerous craters. Using high-resolution images, we identify more than 70 craters (see Crater on Helene in supporting information). Interestingly, Helene shows a bimodal appearance—the heavily cratered trailing hemisphere exhibits a crater density ten times greater than the smooth-looking leading hemisphere (Figure 1). In the case of Enceladus, the deficiency of craters has been suggested to result from a combination of viscous relaxation and burial of craters by both the south polar plume and possibly E-ring material [Kirchoff and Schenk, 2009]. However, a small satellite, such as Helene, is not disturbed by endogenic activity, which provides an optimal situation to examine its possible interactions with E-ring materials.

We study 437 high-resolution images of Helene (500 m/pixel or better) obtained by the Cassini spacecraft through 7 flybys between 2006 and 2011 and find that Helene has numerous, sharply curved gully-like depressions (hereafter streaky depressions; Figure 2). Streaky depressions exist on slopes on the leading hemisphere typically as a group exhibiting similar orientations. These streaky depressions on the leading hemisphere can be identified on images whose resolutions are as high as 200 m/pixel; however, images of the trailing hemisphere, whose resolutions are better than 200 m/pixel, do not show similar features. This indicates that no streaky depressions exist on the trailing hemisphere (Figure 2d).

We construct a numerical shape model of Helene to calculate the local gravitational gradients (Figure 1; see Numerical shape model in supporting information) and critically compare these gradients to the distribution of streaky features (Figure 3). We find that streaky depressions exist only on slopes and strictly follow the local gravitational slopes (Figure 3), which indicates that the streaky depressions result from gravity-induced mass movements. Terrestrial gullies sometimes show structures—such as alcoves (funnelform depressions extending from the top of the slope), channels (linear depressions extending from the narrow stem of an alcove), and fans (cone-shaped deposits crossed by streams)—that indicate transport of material from the top to bottom of the slopes [McClung and Schaerer, 1993]. Analogously to such terrestrial gullies, streaky depressions on Helene sometimes exhibit alcoves and channels (Figures 2b and 2c), further supporting the idea that streaky depressions are formed by gravity-induced mass movements.

The slope angles where streaky depressions exist generally range from $7 \pm 3$ degrees to 20 degrees, which is significantly smaller than the friction angles of terrestrial rock materials (typically, 25–30 degrees or larger [Lambe and Whitman, 1979]). This observation indicates that streaky depressions form in material which collapses easily and exhibits a small friction coefficient, such as fine particulate material rather than massive ice blocks. Perhaps, such fine particles continuously collapse and form the streaky depressions. We conclude that the leading hemisphere of Helene is generally covered by such fine particles because (i) the leading hemisphere generally appears smooth as evidenced by the lack of shadows, even in areas with large illumination angles; (ii) almost no small craters can be identified on the leading hemisphere; and (iii) large craters exhibit flattened shapes. Together, these lines of evidence indicate that deposition of fine particles has modified or erased craters on the leading hemisphere.

3 Discussion

3.1 Plumes From Enceladus Feed Small Satellites

Spectral analysis [Filacchione et al., 2013] shows that the color of Helene is similar to that of Enceladus. Also, in orbits far beyond Enceladus, satellites are expected to overtake E-ring particles (i.e., Enceladus-derived particles), which would result in preferential deposition on the leading hemisphere of Helene. Moreover, E-ring particles are known to be fine particles [Kargel, 2006], consistent with the apparent fine particulate nature of deposits on Helene's leading hemisphere. These facts indicate that the deposit on Helene comes from the E-ring.

Our findings indicate that ice particles ejected from Enceladus are preferentially deposited on Helene's leading hemisphere, and that the resulting deposits occasionally collapse to form streaky depressions. Motivated by this finding, we identified similar deposits on other small satellites in the E-ring region, such as Telesto and Calypso (Tethys' Trojan satellites), Pallene, and Methone (Figure 2 and S1 in supporting information), where higher brightness (i.e., presumably higher densities) of the ring is reported [Verbiscer et al., 2007]. For example, craters on Telesto and Calypso generally exhibit softened, blanketed morphologies with indistinct rims (and sometimes such craters are almost entirely erased), similar to craters on the leading hemisphere of Helene. In fact, even large (>5 km diameter) craters on Telesto and Calypso appear to be buried just as small craters are on Helene. Also, Calypso exhibits streaky depressions (Figure 2e), which are similar to those on Helene. Spectral observations demonstrate the contribution of E-ring material on their surfaces [Buratti et al., 2010], which supports the view that, as with Helene, E-ring material accumulated on these satellites into thick deposits. Moreover, high-resolution images of Pallene (Figure 2g), one of the three Alkyonides satellites, indicate a circular outline with a sharp terminator line lacking any undulations even at the highest resolution of about 500 m/pixel. This implies that Pallene has a featureless spherical shape, which is unusual for a body of only ~2.2 km radius. Methone (Figure 2h), one of the other two Alkyonides satellites, is likely quite similar to Pallene in terms of size, shape, and smooth appearance [e.g., Thomas et al., 2013]. The shapes of these bodies are also possibly related to the accumulation of E-ring material on their surfaces, as such material can cover the original irregular topography.

### 3.2 Lack of Hemispheric Dichotomy for Telesto/Calypso

Unlike Helene, Telesto and Calypso appear to have smooth surfaces and few small craters even on their trailing hemispheres (though Calypso's trailing hemisphere has yet to be observed at high resolution), which suggests that Telesto and Calypso lack any dichotomy (see Figure S1 in supporting information). This diversity may be explained by the motion of the E-ring particles.

The mean velocity difference of a particle ejected from Enceladus relative to an encountered satellite will depend on distance from Saturn. This would naturally cause the impact distribution of particles onto the satellite surface to vary with distance from Saturn as well (see Lack of hemispheric dichotomy for Telesto/Calypso in supporting information). (i) Inside the orbit of Enceladus, particles generally move faster than encountered satellites, which results in deposition on the trailing hemisphere, as is observed for Mimas. (ii) In the orbit immediately beyond Enceladus, the mean relative velocity is modest but non-zero. Particles will move both faster and slower than encountered moons, but on average will be slightly slower. This can explain the small but not negligible albedo difference (1.1 times) between the leading and trailing hemisphere of Tethys (a co-orbital moon of Telesto and Calypso). Still, particles will encounter both the leading and trailing hemisphere of encountered satellites, and this can explain the deposition everywhere on Telesto and Calypso. (iii) Finally, in orbits far beyond Enceladus, the mean relative velocities may be large, with satellites rapidly overtaking Enceladus-derived particles, which results in preferential deposition on the leading hemisphere of Helene, Dione, and Rhea. We note that bombardment by corotational plasma or energetic electrons also influences the color and albedo pattern on mid-sized E-ring satellites [Schenk et al., 2011].

The lack of a hemispheric dichotomy on Telesto and Calypso may be due to not only E-ring particle dynamics but also to possible non-synchronous rotation of these satellites. Non-synchronous rotation re-orients the satellite, causing the leading and trailing hemispheres to migrate across the satellite figure and preventing preferential particle deposition on any specific hemisphere of the satellite. In addition to Telesto and Calypso, Pallene and Methone also appear to lack any hemispheric dichotomy. This may likewise be explained by E-ring particle dynamics or non-synchronous rotation; however, the depositional mechanics on Pallene or Methone may be more complex (see Deposition of ring particles on Pallene and Methone in supporting information).

### 3.3 Thickness of E-Ring Deposits

Shadows of streaky depressions indicate that their depths are typically a few tens of meters, which implies that the particle deposits in which the depressions form must be at least tens of meters thick. Moreover, the leading hemisphere shows the near-absence of small craters (less than ~3 km across; Figure 1c). This indicates that the deposits are unlikely to be more than a few hundred meters thick. Therefore, we estimate that the thickness of the deposits on Helene accumulated from the E-ring is between 10 and 300 m.

On the other hand, Telesto and Calypso almost completely lack large (>5 km diameter) craters. Moreover, the few existing craters are almost entirely erased. This fact suggests that E-ring deposits on Telesto and Calypso are roughly twice as deep as those on Helene. This difference may be due to the densities of E-ring. The E-ring is known to be denser closer to Enceladus [Verbiscer et al., 2007]. Therefore, the E-ring at the orbit of Telesto/Calypso exhibits higher brightness than at the orbit of Helene, which may explain the thicker deposits on Telesto/Calypso relative to those on Helene.

Unlike the deposits on these small satellites, the E-ring deposits on Tethys and Dione are probably quite thin because (1) these satellites' radar-optical albedo appears to decrease with distance from Enceladus [Ostro et al., 2010]; (2) crater statistics [Kirchoff and Schenk, 2010] indicate no deficiency of craters on either Tethys or Dione, in contrast to Helene and Telesto; and (3) high-resolution images of Tethys and Dione show no unambiguous evidence for thick deposits, such as streaky depressions. Nevertheless, albedo [Verbiscer et al., 2007], spectral [Filacchione et al., 2010], and thermal inertia [Howett et al., 2010] measurements indicate thin but non-zero deposits of E-ring particles on these mid-size satellites. Thus, the E-ring particles are likely deposited widely on Tethys and Dione, but the resulting deposits are much thinner than those on the small satellites. The reasons for this difference are unknown. Possible explanations are differences in the dynamics of E-ring particles among the satellites or higher impact velocity onto the mid-sized satellites.

**4 Implication for Ephemeral Cryovolcanism on Enceladus**

We perform an age estimate of the E-ring deposits based on the cratering rate. The high crater density on the trailing hemisphere of Helene indicates that Helene is basically an old object. Based on the crater chronology of the saturnian system [Zahnle et al., 2003], we estimate that the surface age of the heavily cratered terrain on the trailing hemisphere is ~4.0 Gy (and at least ~1.0 Gy), which coincides with the age estimated from the distribution of large (>5 km) craters (Figure 1c). We also find that the distributions of craters exceeding 5 km in diameter on Helene and Telesto are generally similar to those of Dione and Tethys [Kirchoff and Schenk, 2010] (Figure 1c), which indicates that, in general, the original crater densities of small satellites are similar for these mid-sized satellites. Thus, Helene, Telesto, and Calypso probably have similar formational ages, which are significantly older than that of the E-ring deposits.

On Helene's E-ring deposit, whose area is 1637 km² estimated from our shape model, we identify five craters of ~200 m diameter but no craters exceeding 1 km in diameter. If we adopt standard cratering-rate estimates for the outer solar system [Zahnle et al., 2003], the best estimates for deposit ages are several tens of My or younger (see Age estimates based on cratering rate in supporting information). The crater size-frequency distribution of the trailing hemisphere of Telesto is similar to that of the leading hemisphere of Helene.

Interestingly, several lines of evidence suggest that Enceladus has not been active at its current level over solar-system history. For example, Kargel [2006] shows that no geological evidence exists to support a large change in its radius, which would be expected if the current mass-loss rates have been maintained through Enceladus' lifetime. Moreover, Roberts and Nimmo [2008] show that the current energy output of Enceladus is difficult to sustain over solar-system history. Enceladus' current heat flux greatly exceeds that occurring in equilibrium with its current eccentricity given the time-averaged Q of Saturn [Meyer and Wisdom, 2007], suggesting that the current activity may be greater than average. Recent global geophysical models show that Enceladus can exhibit episodic activity, with short periods of intense activity interspersed with long quiescent epochs [Showman et al., 2013]. These studies are consistent with our views. Thus, there is the possibility that the accumulation of E-ring material on small satellites in this region began several My ago as a result of the initiation of cryovolcanism of Enceladus.

## Acknowledgments

This work is supported in part by Grant-in-Aid for JSPS Fellows (to NH), JSPS KAKENHI (to HM), and the NASA Origins program (to APS). We use the raw data freely available via NASA's Planetary Data System (http://pds.nasa.gov/).

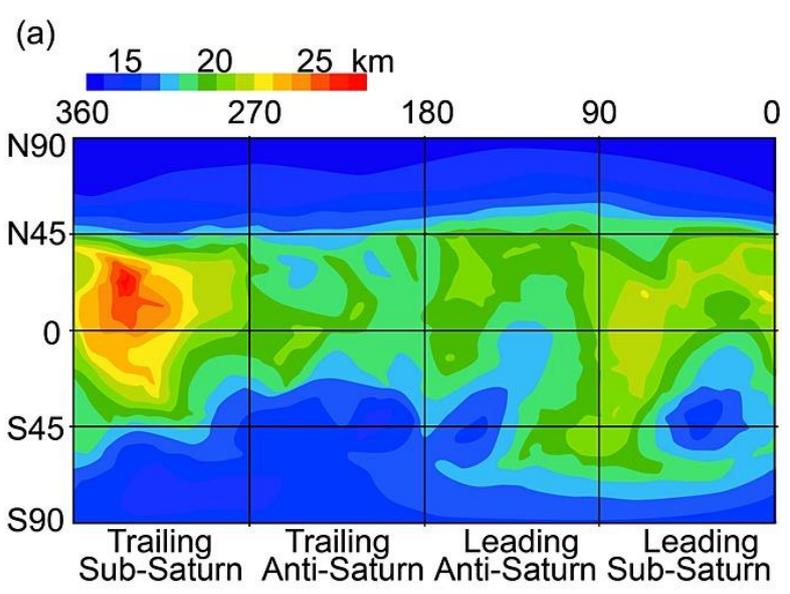

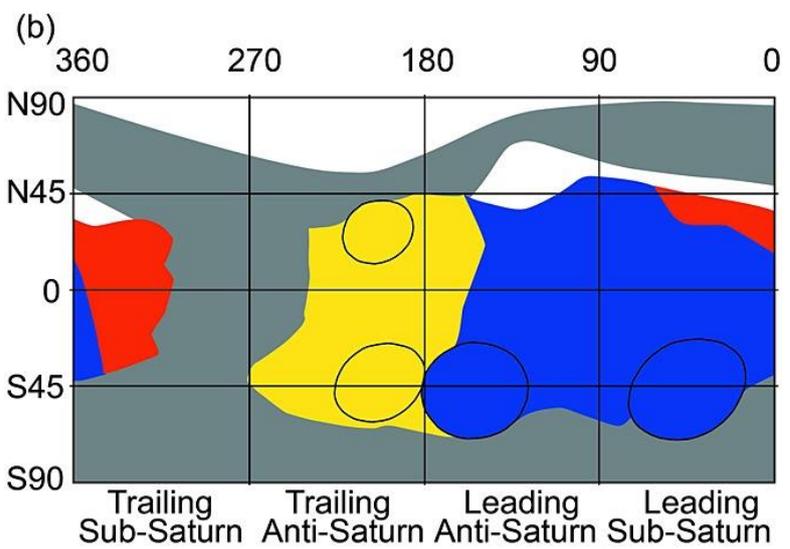

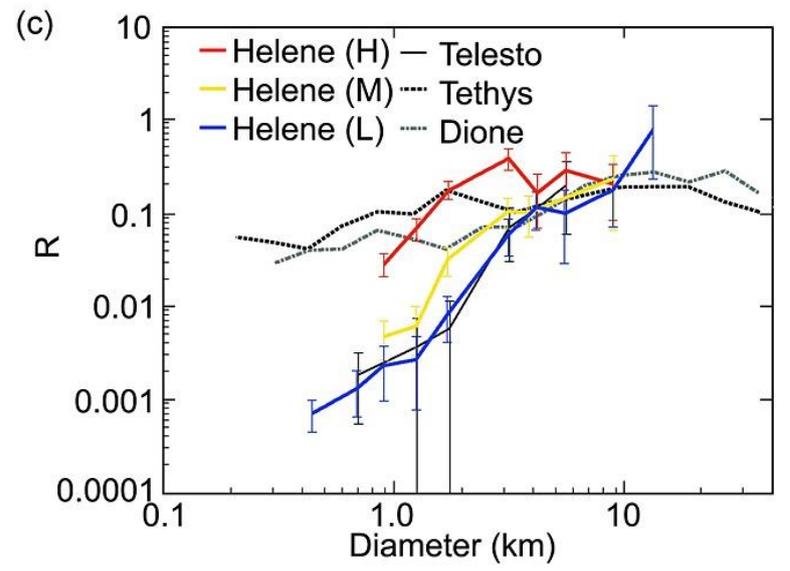

**Figure 1** Topography and crater distributions of Helene. (a) Global topographic map of Helene with colors representing topographic height relative to the geometric center. (b) Regional map with heavily cratered terrain (red), moderately cratered terrain (yellow), less cratered terrain (blue), features unidentified (gray), and no images obtained (white). Black circles indicate large craters (~10 km). (c) Relative size-frequency distribution of craters of heavily cratered terrain (H), moderately cratered terrain (M), and less cratered terrain (L) of Helene as well as Telesto, Tethys, and Dione. The data for Tethys and Dione come from Kirchoff and Schenk [2010]. Colors represent regions in Figure 1b.

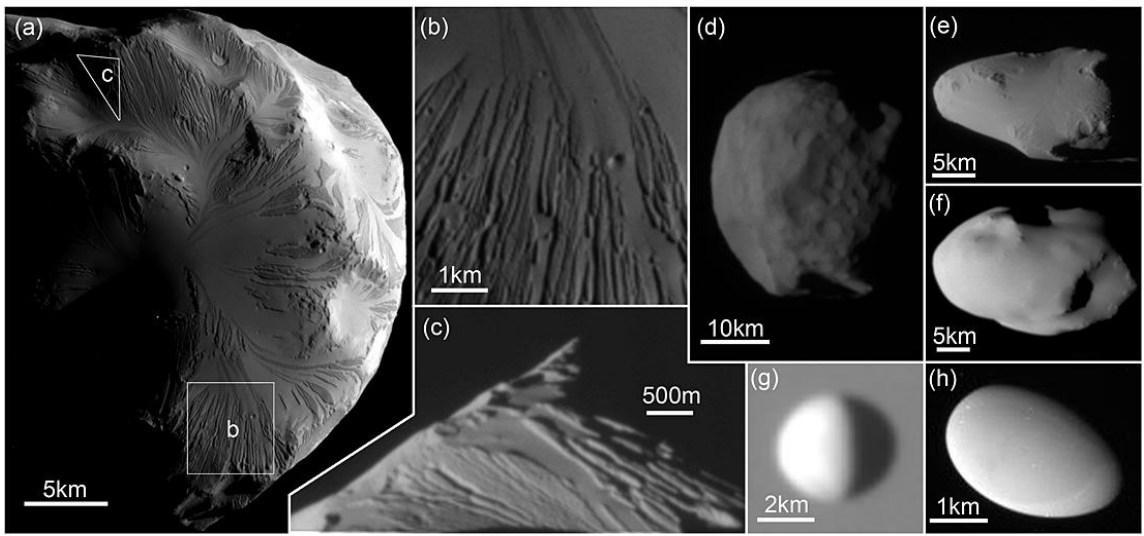

**Figure 2** Small satellites in the E-ring region. (a) Helene's leading hemisphere (image N1687119539). Insets indicate locations of Figures 2b and 2c. (b) Close-up image of the streaky depressions (N1687119539; 42 m/pixel). (c) The highest-resolution image of Helene (N1646317865; 24 m/pixel). The depths of the streaky features are estimated to be a few tens of meters. (d) Helene's trailing hemisphere (N1563643679). (e) Calypso's leading hemisphere (N1644754662). (f) Telesto's leading hemisphere (N1630076968). (g) Anti-Saturn side of Pallene with Saturn as a background (N1665947247; 223 m/pixel). (h) Methone's leading hemisphere (N00189072).

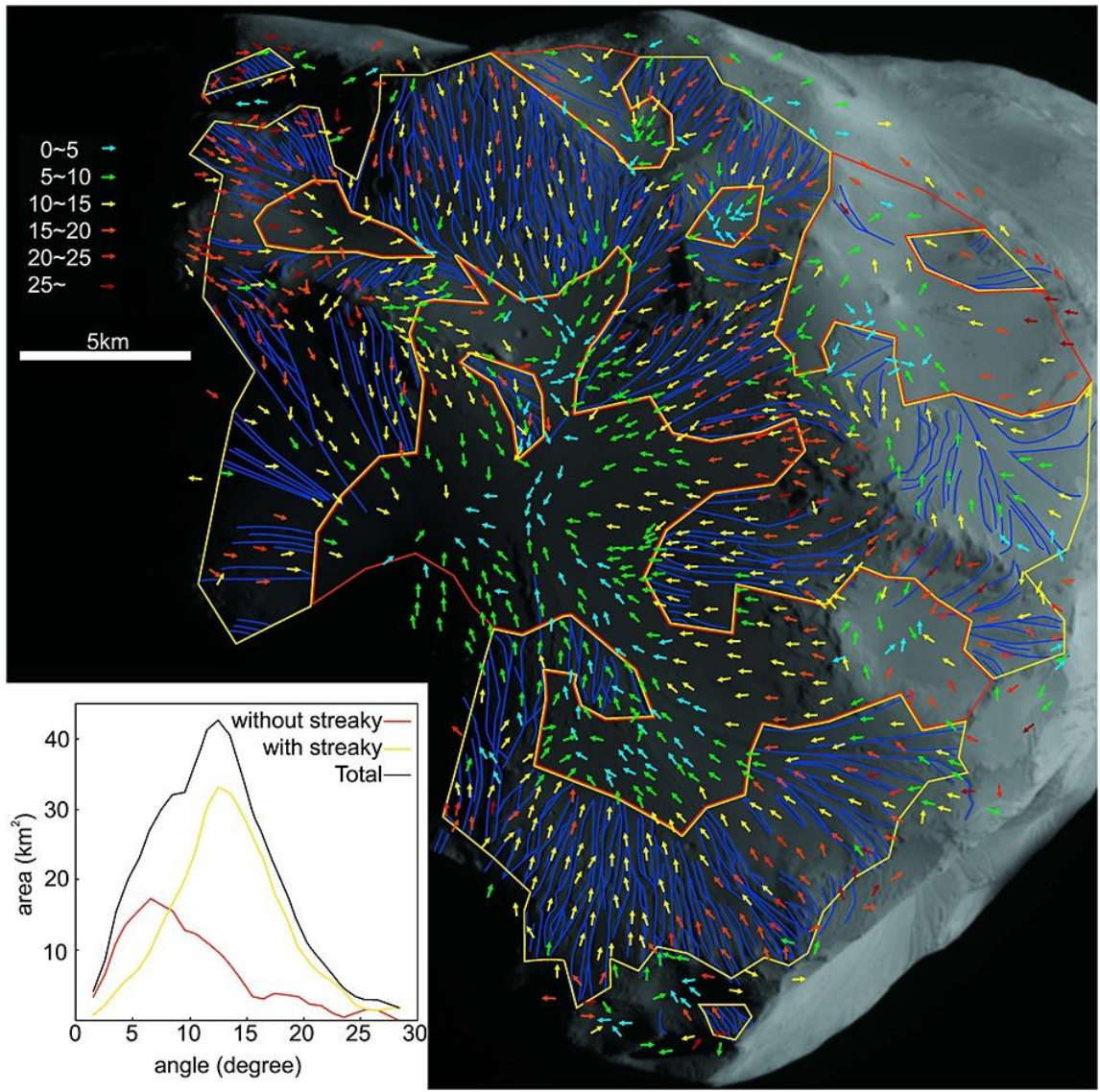

Figure 3 The leading hemisphere of Helene (N1687120437) with arrows (located at the centers of triangles of the shape-model) indicating the directions of the local gravity (color represents slopes). Blue lines indicate streaky depressions. The regions enclosed by yellow or red lines are the regions with and without streaky depressions, respectively. Streaky features exist on steeper (> ~7 degree) and strictly follow the directions of surface gravity. Inset shows the total areas of both regions as a function of slope angle (see Table S3 in supporting information). The horizontal axis is a moving average of ±1.5 degree for the total area of the angle ±0.5 degree.

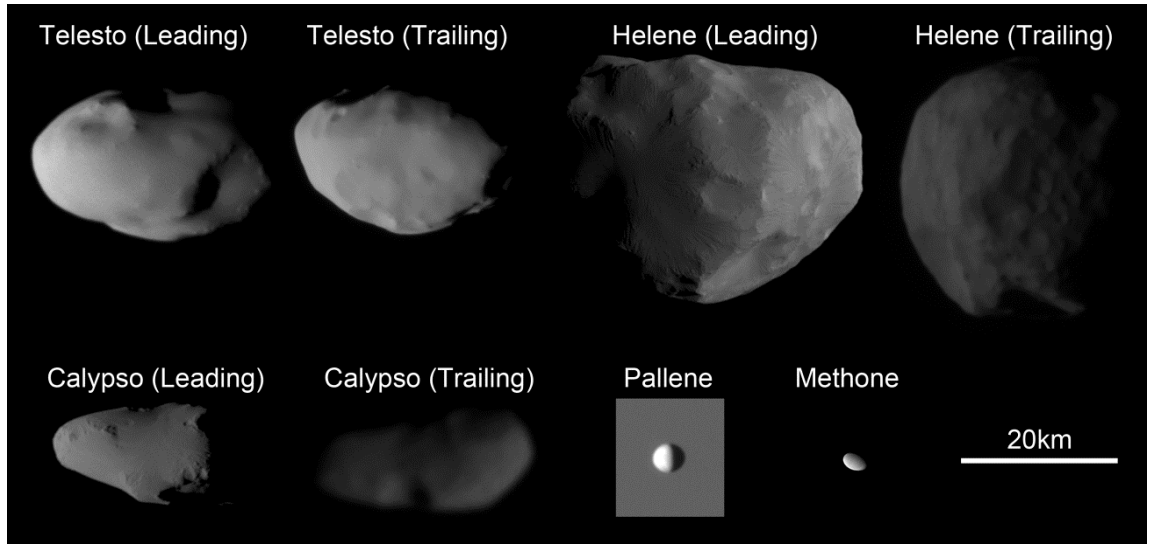

**Figure S1.** Small satellites in the E-ring region shown in the same scale (Cassini images N168712110 and N1563643679 for Helene; N1630076968 and N1514163666 for Telesto; N1644754662 and N1506184171 for Calypso; N1665947247 for Pallene; and N00189072 for Methone). Leading and Trailing indicate the leading and trailing hemispheres, respectively.

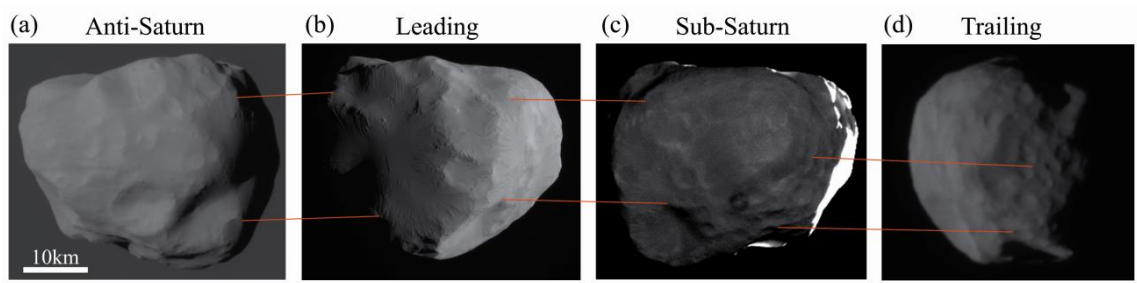

**Figure S2.** Images used for constructing the shape model (not all of them; see Table S1). Red lines connect the same signature points between images. (Image numbers from left to right, N1646319549, N1687121104, N1646315085, and N1563643679).

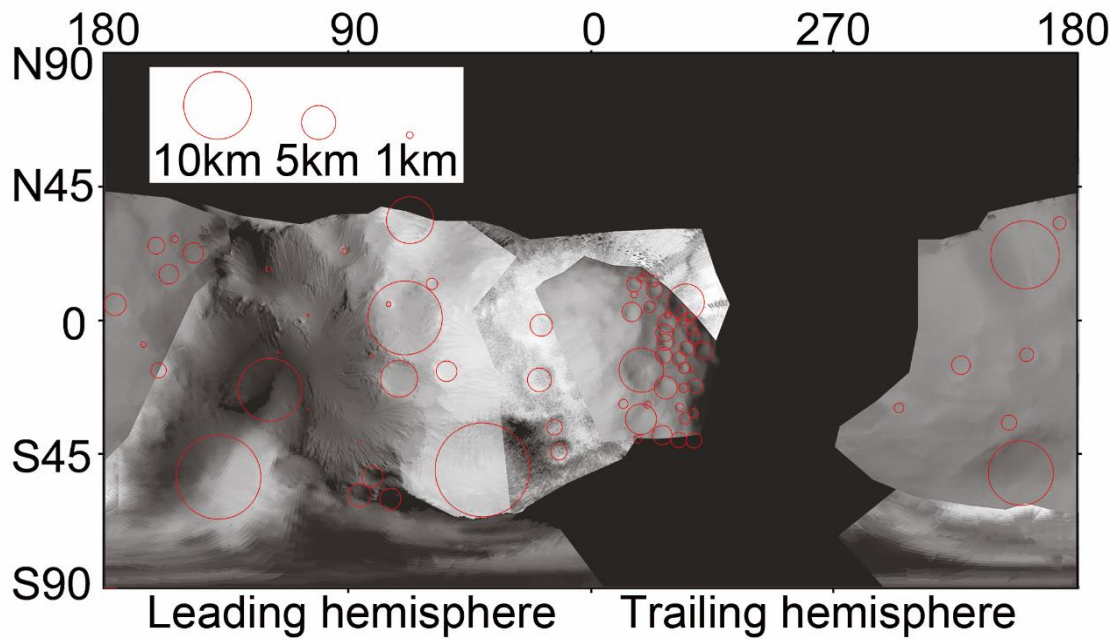

**Figure S3.** The distribution of craters (red circles) on Helene on a cylindrically-projected global image created from images listed in Table S1. Black region lacks images (craters cannot be identified in this region). Note that we identified craters in raw images rather than this projected image. Some craters on the anti-Saturn side of the trailing hemisphere (longitude 180-270) might remain unidentified due to low solar incidence angles of the images.

## Supplementary Information

### Crater on Helene

As observed on most satellites of Saturn, Helene has numerous craters. Using the high-resolution images, we identify more than 70 craters. Craters are identified by (i) a circular depression with a rim, (ii) a depression imaged with a resolution of at least 5 pixels for clear understanding of its shape, and (iii) a depression identified in more than one image. Larger craters (> 10 km) on north and south polar terrain are also verified by our shape model. Craters are not uniformly distributed over the surface of Helene (Fig. S3). At least 5 craters are larger than ~10 km in diameter and at least 8 craters are 5 to 10 km in diameter. They are almost uniformly distributed over the entire surface of Helene. In contrast, the number of small craters (below 5 km in diameter) varies significantly depending on the region. We identify three distinct regions based on the crater densities, such as (i) heavily cratered terrain (red in Fig. 1b), which covers the sub-Saturn side of the trailing hemisphere (we identify 38 craters ranging from 0.7 km to 10 km in diameter), (ii) moderately cratered terrain (yellow in Fig. 1b), which covers the anti-Saturn side of the trailing hemisphere (14 craters ranging from 0.2 km to 12 km), and (iii) less cratered terrain (blue in Fig. 1b), which covers the leading hemisphere (22 craters ranging from 1 km to 10 km). We note that crater identifications in the moderately cratered terrain suffered from low solar-incidence angles in most images of this region. The locations and distributions of identified craters are shown in Fig. S3. Fig. 1c shows crater size-frequency distributions in a relative plot (R-plot; the ratio of the differential form of cumulative crater-size distribution to a size-frequency distribution with differential slope equal to 3. Error bars are defined as $\pm R*N^{-0.5}$, where N is the number of craters between a bin. Similarly, we analyze the Telesto's trailing hemisphere to obtain the R-plot, which is shown in Fig. 1c. For these analyses, we follow the method of Crater Analysis Technique Working Group [1979].

### Numerical shape model

To develop a numerical shape model, we follow the previously proposed method [*Hartley and Zisserman*, 2003] by using 16 high-resolution images shown in Table S1 to develop a shape model. We i) split the surface of Helene into 10 parts; ii) select 3 images from the 16 high-resolution images showing a particular part of the body; iii) identify at least 8 corresponding points in all of these 3 images; and iv) measure both relative locations of these corresponding points and camera positions of 3 images using the epipolar geometry, which results in forming a local shape model (Fig. S2). We perform the process of ii) to iv) for all 10 parts of Helene before binding the local shape models into a single very-rough shape model (Table S2).

Once a shape model is developed, a camera position for each image can be precisely determined. We then perform stereo analyses for all of the 16 high-resolution images with the camera positions determined to obtain the exact locations of 750 control points, which results in quality improvement of the shape model. We compare the shape model to all of the high-resolution images obtained by the Cassini spacecraft to confirm that the shape model properly reproduces outlines and features in these images. In particular, the leading hemisphere is carefully analyzed by defining 1024 polygons to critically compare the local gravity and the directions of streaky depressions.

We develop a topographic map of Helene based on the shape model assuming that the center of mass coincides with the geometric center (Fig. 1a). The map enables us to identify the existence of basins, which are divided by ridges. We assume a constant density throughout the body, which might not be realistic for deriving the exact values of gravity but is adequate for obtaining approximate gravity-vector orientations and determination of downslope directions.

We constructed the numerical shape model as the assemblage of 7492 small tetrahedrons, whose densities were assumed to be the same. The leading hemisphere is specifically analyzed carefully by defining 1024 polygons. To derive the local gravity, we i) measure the distance between the geometric center of a single tetrahedron at the centroid of a given polygon; ii) calculate the gravitational vector from the distance and the volume of the tetrahedron; iii) perform i) and ii) for all tetrahedrons consisting of the shape model; iv) integrate these gravitational vectors obtained by iii) to obtain the total gravity acceleration from entire Helene at the given polygon; v) calculate the inner product between the normal vector of the polygon and the total gravitational vector on the polygon to obtain the angle of the slope; vi) remove the component along the normal vector of the polygon from the gravitational vector to obtain the direction of the local gravity on the polygon; vii) perform i) to vi) for all 1024 polygons consisting of the surface of leading hemisphere.

**Lack of hemispheric dichotomy for Telesto/Calypso**

We consider that the lack of hemispheric dichotomy for Telesto/Calypso can be explained by either non-synchronous rotation or the motion of the E-ring particles. The latter case is discussed below.

The dynamics of the E-ring particles are difficult to directly observe and can be surprisingly complex due to the competing effects of the gravity of satellites, solar radiation pressure, electrostatic grain potential, the Lorenz forces, the plasma-sputtering, and plasma drag [*Horányi et al.*, 2008, 2009]. Nevertheless, these forces cause large orbital eccentricities or an increasing of semi-major axes, which result in correspondingly large radial excursions of E-ring particles in just a few years [*Horányi et al.*, 1992, 2009]. Especially, large orbital eccentricities of E-ring particles having the semi-major axis close to Enceladus can play an important role in the hemispheric dichotomies on satellites. Within the orbit of Enceladus, E-ring particles move faster than satellites, which causes E-ring particles to collide preferentially onto the trailing side of Mimas. On the other hand, beyond the orbit of Enceladus, E-ring particles move slower and are overtaken by satellites, which causes E-ring particles to collide preferentially onto the leading side of Dione and Tethys. This can also explain Helene's hemispheric dichotomy.

Unlike Helene, Telesto and Calypso only faintly show a hemispheric dichotomy (e.g., Telesto's leading hemisphere has a relatively smoother and relatively less cratered surface than the trailing side). This, again, could be explained by the differences between the relative velocities of E-ring particles relative to Dione and Tethys. Because Tethys is closer to Enceladus than Dione, the velocities of E-ring particles in the orbit of Tethys is less decelerated compared with those in the orbit of Dione. Therefore, interactions with particles might not be concentrated significantly on the leading hemisphere of Tethys, Telesto, and Calypso.

**Deposition of ring particles on Pallene and Methone**

Spherical shapes and smooth appearances of both Pallene and Methone might also be related to the dense E-ring particles at their orbits but might be complicated by the existence of the co-orbital ring and the arc coexisting with both Pallene and Methone (perhaps also Anthe, the other of the Alkyonides), whose materials might originate from these satellites [*Hedman et al.*, 2009]. In any case, both of these tiny satellites are difficult to immediately accumulate materials released by impacts, which may temporary form the ring or arcs. Nevertheless, these released materials, originally from both the E-ring and satellites, should re-accumulate on the satellites because of the confinement of dust in resonances with Mimas or Enceladus [*Hedman et al.*, 2009]. These processes may explain their unusual smooth surfaces and spherical shapes. We suspect that Anthe also has a spherical shape with a smooth appearance.

**Age estimates based on cratering rate**

Because of the nature of crater chronology, the crater densities on Enceladus do not directly indicate when Enceladus' cryovolcanism has begun (the crater densities on Enceladus may show a state of dynamic equilibrium between the cratering and resurfacing rates). The situation on Helene is different from that of Enceladus; particles from Enceladus may deposit over the surface of Helene but its thickness should be fairly thin (a few ten or hundred meters), while the depth of a crater is relatively deep enough (a crater with a few hundred meters diameter should have a few tens of meters depth). In other words, the E-ring depositions are difficult to bury newly-formed craters (larger than ~1 km in diameter) after an initiation of the deposition. Thus, we consider that the deposits on Helene are suitable to discuss when the Enceladus' cryovolcanism has begun.

On Helene's E-ring deposit, whose area is 1,637 km², we identify 5 craters of ~200 m in diameter but not exceeding 1 km. This clearly indicates young formational ages of the deposits, but estimating the ages can be more complicated than the cases on the Moon. This is mostly due to the lack of dated samples from the saturnian satellites and to the difficulty in determining the cratering rate, which should be estimated theoretically based on the populations and orbits of potential impactors. According to *Dones et al.* [2009], potential impactors come from (i) main-belt asteroids, (ii) trojan objects of the gas giants, (iii) Centaurs and ecliptic comets, (iv) Saturn's irregular satellites, (v) planetocentric bodies, and (vi) Nearly Isotropic Comets.

Previous theoretical studies for estimating formational ages of geological features on saturnian satellites [e.g. *Zahnle et al.*, 2003] are based either on (A) the crater densities on the Moon, Europa, Ganymede, and Triton, and (B) the statistics of encounter with potential impactors, such as comets and asteroids. Trans-Saturnian objects, such as Centaurs, are considered as the major impactors. Their populations and orbital properties [*Gladman et al.*, 2001] are studied with ground-based observations, even though the accurate size-frequency of small objects is still unknown because of their great distances. *Zahnle et al.* [2003] estimates the cratering rate for the outer solar system, including that on saturnian small satellites; for example, the cratering rate for objects larger than 1 km in diameter on Helene is (A) $2.0 \times 10^{-7}$ craters per $10^3$ km$^2$ per year if we assume that small objects obey a nearly collisional distribution and (B) $2.0 \times 10^{-9}$ craters per $10^3$ km$^2$ per year if we assume the size–number distribution is like that inferred at Jupiter. Therefore, assuming this cratering rate, the formational ages of the E-ring deposits on Helene are likely to be 0.3-5 My (the former case) or 200-500 My (the latter case).

On the other hand, according to *Dones et al.* [2009] if major potential impactors originated from heliocentric objects, such as Centaurs, impact cratering on saturnian satellites should exhibit a hemispheric dichotomy because heliocentric impactors strongly favor cratering of the leading hemisphere. However, small craters on saturnian satellites do not show such dichotomies, which may imply that planetocentric impactors, such as secondary impactors, are dominant to form smaller craters. If we assume that the heavily cratered regions on Helene, which have 100 craters larger than 1km in diameter per 1000 km$^2$, was formed more than 4 Gy and that the cratering rate is constant for its life time, we obtain ~40 My for the formational age of the E-ring deposit on Helene. Or, if we assume the crater distributions of Dione [*Kirchoff and Schenk*, 2010], whose Trojan satellites include Helene, indicate that the crater formation rates of > 1km craters is $7.5 \times 10^{-6}$ craters per $10^3$ km$^2$ per year and of > 200 m craters is ~$10^{-7}$ craters per $10^3$ km$^2$ per year, we obtain ~50 My for the formational age of the E-ring deposit on Helene.

We note that these estimates include ambiguities in their assumptions. However, importantly, all these independent estimates coincide around several 10 My. Thus, we conclude several tens of million years or younger for the most likely formational ages of E-ring deposit on Helene.

Table S1. Images used for analyses of a shape model of Helene.

| Image | Date (UTC) | Range (km) | Resolution (m/pixel) | Sub-Cassini Lat. | Sub-Cassini Lon. | Sub-Solar Lat. | Sub-Solar Lon. |
|---|---|---|---|---|---|---|---|
| N1519536732 | 2006-2-25 | 67836 | 423.7 | -0.4 | 96.6 | -18.5 | 187.2 |
| N1519537272 | 2006-2-25 | 68271 | 426.4 | -0.4 | 93.6 | -18.5 | 188.0 |
| N1534480072 | 2006-8-27 | 50353 | 314.5 | 73.8 | 31.0 | -16.2 | 237.9 |
| N1534483492 | 2006-8-27 | 61949 | 387.0 | 75.4 | 138.9 | -16.2 | 243.0 |
| N1563643679 | 2007-7-20 | 38890 | 242.9 | -2.5 | 316.2 | -11.4 | 13.4 |
| N1563644326 | 2007-7-20 | 38466 | 240.3 | -2.6 | 321.6 | -11.4 | 14.4 |
| N1506207298 | 2008-11-24 | 69067 | 431.4 | 25.7 | 347.3 | -3.9 | 329.6 |
| N1646315085 | 2010-3-3 | 22073 | 137.9 | -3.8 | 4.0 | 3.2 | 193.1 |
| N1646319549 | 2010-3-3 | 18821 | 117.6 | -3.6 | 183.0 | 3.2 | 199.9 |
| N1646320608 | 2010-3-3 | 28482 | 177.9 | -2.2 | 186.0 | 3.2 | 201.5 |
| W1646317554 | 2010-3-3 | 1911 | 119.4 | -42.4 | 116.4 | 3.2 | 196.9 |
| W1646317899 | 2010-3-3 | 4113 | 256.9 | -18.2 | 164.8 | 3.2 | 197.1 |
| N1675165048 | 2011-1-31 | 31434 | 190.3 | -3.5 | 114.3 | 8.2 | 173.3 |
| N1687119135 | 2011-6-18 | 7355 | 44.5 | 2.7 | 147.3 | 10.1 | 19.2 |
| N1687121104 | 2011-6-18 | 9800 | 59.3 | 1.3 | 87.4 | 10.1 | 22.2 |
| N1687121464 | 2011-6-18 | 11031 | 66.8 | 1.0 | 81.8 | 10.1 | 22.8 |

Table S2. The distance (in meter) between the geometric center and the surface based on the shape model of Helene (latitudes in horizontal lines and longitudes in vertical lines)

|     | -90   | -80   | -70   | -60   | -50   | -40   | -30   | -20   | -10   | 0     |
|-----|-------|-------|-------|-------|-------|-------|-------|-------|-------|-------|
| 10  | 15719 | 16069 | 17066 | 17412 | 17580 | 17316 | 17762 | 18518 | 18627 | 19301 |
| 20  | 15719 | 16118 | 17164 | 17965 | 17136 | 16887 | 16965 | 17477 | 17615 | 19290 |
| 30  | 15719 | 16788 | 17667 | 17833 | 16614 | 15878 | 16488 | 17165 | 17942 | 19555 |
| 40  | 15719 | 16832 | 18213 | 17387 | 16331 | 15584 | 16611 | 17680 | 18183 | 19842 |
| 50  | 15719 | 17436 | 18514 | 18063 | 16891 | 16483 | 17713 | 19066 | 19259 | 20455 |
| 60  | 15719 | 17083 | 18489 | 18551 | 17935 | 17599 | 19735 | 20206 | 21084 | 21303 |
| 70  | 15719 | 16629 | 18600 | 19465 | 18868 | 19917 | 21033 | 21439 | 21834 | 21483 |
| 80  | 15719 | 16860 | 18305 | 19950 | 20168 | 20823 | 20662 | 20580 | 21029 | 21519 |
| 90  | 15719 | 16642 | 18014 | 19834 | 20371 | 19991 | 19891 | 19973 | 20445 | 20190 |
| 100 | 15719 | 17323 | 17578 | 19500 | 20204 | 19788 | 19120 | 19000 | 19083 | 19171 |
| 110 | 15719 | 16775 | 17773 | 18689 | 19949 | 19526 | 18859 | 18368 | 18413 | 18304 |
| 120 | 15719 | 16493 | 17154 | 18547 | 19206 | 19136 | 18405 | 17958 | 17485 | 17800 |
| 130 | 15719 | 16950 | 17247 | 19127 | 18399 | 18584 | 17652 | 17296 | 17209 | 17786 |
| 140 | 15719 | 16684 | 17029 | 18054 | 17497 | 16505 | 16472 | 17584 | 17829 | 18238 |
| 150 | 15719 | 16422 | 17193 | 17202 | 15845 | 15814 | 16503 | 17854 | 18791 | 18907 |
| 160 | 15719 | 16620 | 17375 | 16870 | 15976 | 16579 | 17272 | 18618 | 19780 | 19309 |
| 170 | 15719 | 16422 | 16842 | 17088 | 16409 | 17156 | 18150 | 19477 | 19448 | 19366 |
| 180 | 15719 | 16471 | 16842 | 16941 | 17107 | 17454 | 17799 | 18700 | 18445 | 18991 |
| 190 | 15719 | 16169 | 16379 | 16599 | 15755 | 15712 | 16400 | 17789 | 18065 | 18734 |
| 200 | 15719 | 15829 | 16120 | 16469 | 15041 | 14804 | 16174 | 17994 | 18074 | 18438 |
| 210 | 15719 | 15892 | 15939 | 16302 | 14986 | 14933 | 16598 | 17877 | 18345 | 19109 |
| 220 | 15719 | 15463 | 15705 | 16039 | 15444 | 16170 | 16367 | 17738 | 17904 | 19962 |
| 230 | 15719 | 15426 | 15421 | 15729 | 15655 | 15901 | 16347 | 17168 | 18645 | 19985 |
| 240 | 15719 | 15245 | 15316 | 15385 | 15476 | 15630 | 16243 | 18403 | 19761 | 20758 |
| 250 | 15719 | 15033 | 15019 | 15713 | 15213 | 15810 | 17398 | 19787 | 20155 | 19799 |
| 260 | 15719 | 14827 | 15143 | 15834 | 15317 | 15910 | 17476 | 18864 | 19565 | 19366 |
| 270 | 15719 | 14721 | 14828 | 15979 | 15638 | 15834 | 16764 | 18138 | 19620 | 19654 |
| 280 | 15719 | 14721 | 14764 | 16167 | 16272 | 16289 | 17253 | 17600 | 19152 | 20311 |
| 290 | 15719 | 15081 | 14699 | 16137 | 17002 | 17549 | 18013 | 18626 | 19235 | 20560 |
| 300 | 15719 | 14751 | 15010 | 15727 | 17884 | 18929 | 18927 | 19303 | 19840 | 20999 |
| 310 | 15719 | 14631 | 14698 | 15496 | 16788 | 19341 | 21007 | 20913 | 21313 | 23335 |
| 320 | 15719 | 14689 | 14624 | 15242 | 16385 | 19013 | 21797 | 23464 | 22170 | 23716 |
| 330 | 15719 | 14990 | 15116 | 15254 | 16356 | 18645 | 21024 | 22527 | 22825 | 24493 |
| 340 | 15719 | 15095 | 15299 | 16394 | 17989 | 18658 | 20591 | 22377 | 23696 | 24295 |
| 350 | 15719 | 15495 | 16033 | 17199 | 18424 | 18873 | 19090 | 20513 | 22235 | 23889 |
| 360 | 15719 | 15448 | 16041 | 17513 | 18104 | 17938 | 18441 | 19771 | 20585 | 21219 |

Table S2. Continued

|     | 10    | 20    | 30    | 40    | 50    | 60    | 70    | 80    | 90    |
|-----|-------|-------|-------|-------|-------|-------|-------|-------|-------|
| 10  | 21110 | 21699 | 20990 | 21250 | 17418 | 15180 | 14501 | 14501 | 14123 |
| 20  | 19741 | 20372 | 21452 | 20763 | 18191 | 14735 | 14735 | 14287 | 14123 |
| 30  | 19724 | 20289 | 21407 | 20435 | 17267 | 15683 | 14287 | 13796 | 14123 |
| 40  | 20360 | 20946 | 20984 | 20420 | 18149 | 15075 | 15075 | 13796 | 14123 |
| 50  | 20839 | 21721 | 21270 | 19618 | 17323 | 15790 | 14996 | 14996 | 14123 |
| 60  | 21115 | 20943 | 19210 | 18956 | 17693 | 16577 | 14996 | 14415 | 14123 |
| 70  | 21676 | 21636 | 19871 | 18491 | 18358 | 15799 | 15257 | 14415 | 14123 |
| 80  | 21169 | 20968 | 20128 | 18906 | 18127 | 16118 | 15257 | 14874 | 14123 |
| 90  | 20488 | 20593 | 20619 | 19961 | 18550 | 17017 | 14874 | 14874 | 14123 |
| 100 | 19355 | 19766 | 19965 | 19896 | 18960 | 17602 | 15093 | 14395 | 14123 |
| 110 | 18664 | 18971 | 19565 | 19568 | 18650 | 16792 | 15093 | 14395 | 14123 |
| 120 | 18336 | 18792 | 19290 | 19966 | 18449 | 16860 | 15425 | 15425 | 14123 |
| 130 | 18842 | 18882 | 19463 | 19914 | 18576 | 17229 | 14680 | 14680 | 14123 |
| 140 | 18975 | 19689 | 19493 | 19744 | 18066 | 16745 | 15382 | 14680 | 14123 |
| 150 | 19892 | 20587 | 20282 | 19704 | 18495 | 15688 | 15688 | 14085 | 14123 |
| 160 | 19437 | 20404 | 20255 | 20236 | 18057 | 15872 | 14906 | 14085 | 14123 |
| 170 | 19885 | 19626 | 19090 | 19412 | 18021 | 16207 | 14906 | 14085 | 14123 |
| 180 | 18384 | 18796 | 18067 | 18318 | 17847 | 15978 | 15370 | 14443 | 14123 |
| 190 | 18416 | 18646 | 19504 | 19225 | 17566 | 15339 | 14443 | 14443 | 14123 |
| 200 | 18809 | 18173 | 18182 | 18656 | 17500 | 15912 | 14443 | 14443 | 14123 |
| 210 | 20217 | 18599 | 18611 | 18428 | 16851 | 15108 | 15108 | 13802 | 14123 |
| 220 | 19348 | 18668 | 19876 | 17972 | 16686 | 15618 | 14365 | 13802 | 14123 |
| 230 | 18947 | 18183 | 19082 | 17919 | 16843 | 15618 | 14365 | 13802 | 14123 |
| 240 | 18796 | 18120 | 17904 | 17510 | 16841 | 14891 | 14891 | 13802 | 14123 |
| 250 | 19151 | 18110 | 18039 | 18296 | 15774 | 15774 | 14141 | 14141 | 14123 |
| 260 | 19405 | 18434 | 19003 | 18438 | 16835 | 14970 | 14141 | 14141 | 14123 |
| 270 | 20183 | 19375 | 20207 | 19341 | 16546 | 14970 | 14970 | 14141 | 14123 |
| 280 | 21280 | 21238 | 21525 | 20015 | 16663 | 15683 | 14633 | 13796 | 14123 |
| 290 | 21432 | 21322 | 21044 | 19289 | 17001 | 14633 | 14633 | 13796 | 14123 |
| 300 | 22323 | 21829 | 21497 | 18880 | 16428 | 15365 | 13796 | 13796 | 14123 |
| 310 | 24042 | 23414 | 21414 | 18364 | 16768 | 15365 | 14376 | 13796 | 14123 |
| 320 | 24860 | 24022 | 23296 | 19191 | 15850 | 14376 | 14376 | 14376 | 14123 |
| 330 | 25153 | 26533 | 24608 | 19913 | 17419 | 15850 | 14376 | 13888 | 14123 |
| 340 | 25363 | 25322 | 24404 | 20236 | 16073 | 14820 | 13888 | 13888 | 14123 |
| 350 | 23330 | 22643 | 21239 | 21370 | 17362 | 14820 | 13888 | 13888 | 14123 |
| 360 | 22822 | 21701 | 20832 | 20573 | 18902 | 15577 | 14501 | 13888 | 14123 |

Table S3. Angles of slopes to areas calculated from the shape model of Helene.

| Angle | Area with streaky depressions (km²) | Area without streaky depressions (km²) | Total (km²) |
|---|---|---|---|
| 0 ~ 1 | 0.00 | 2.23 | 2.23 |
| 1 ~ 2 | 0.70 | 2.88 | 3.58 |
| 2 ~ 3 | 1.54 | 4.81 | 6.35 |
| 3 ~ 4 | 3.86 | 11.62 | 15.48 |
| 4 ~ 5 | 7.04 | 16.75 | 23.79 |
| 5 ~ 6 | 6.65 | 13.18 | 19.83 |
| 6 ~ 7 | 8.48 | 16.88 | 25.36 |
| 7 ~ 8 | 14.75 | 21.54 | 36.30 |
| 8 ~ 9 | 17.52 | 10.81 | 28.33 |
| 9 ~ 10 | 17.58 | 14.17 | 31.75 |
| 10 ~ 11 | 23.97 | 13.37 | 37.34 |
| 11 ~ 12 | 31.08 | 9.52 | 40.60 |
| 12 ~ 13 | 35.80 | 10.27 | 46.08 |
| 13 ~ 14 | 32.15 | 9.22 | 41.36 |
| 14 ~ 15 | 29.45 | 4.93 | 34.38 |
| 15 ~ 16 | 27.17 | 2.86 | 30.04 |
| 16 ~ 17 | 21.58 | 2.34 | 23.92 |
| 17 ~ 18 | 19.83 | 3.88 | 23.72 |
| 18 ~ 19 | 12.43 | 5.35 | 17.78 |
| 19 ~ 20 | 11.34 | 1.18 | 12.53 |
| 20 ~ 21 | 7.18 | 3.59 | 10.78 |
| 21 ~ 22 | 6.09 | 1.96 | 8.05 |
| 22 ~ 23 | 6.36 | 0.57 | 6.93 |
| 23 ~ 24 | 3.87 | 0.49 | 4.36 |
| 24 ~ 25 | 2.31 | 0.00 | 2.31 |
| 25 ~ 26 | 0.55 | 2.67 | 3.22 |
| 26 ~ 27 | 1.44 | 1.83 | 3.27 |
| 27 ~ 28 | 2.05 | 0.00 | 2.05 |
| 28 ~ 29 | 1.63 | 0.00 | 1.63 |
| > 29 | 1.07 | 0.00 | 1.07 |